\documentclass[
]{ceurart}

\usepackage{listings}
\usepackage{amsmath,amssymb,amsfonts}
\usepackage{algorithmic}
\usepackage{graphicx}
\usepackage{textcomp}
\usepackage{xcolor}

\usepackage{hyperref}

\begin{document}

\copyrightyear{2023}
\copyrightclause{Copyright for this paper by its authors.
  Use permitted under Creative Commons License Attribution 4.0
  International (CC BY 4.0).}

\conference{ICPM 2023: ICPM Doctoral Consortium and Demo Track 2023,
  October 23--27, 2023, Rome, Italy}

\title{OCEL 2.0 Resources -- www.ocel-standard.org}

\author[]{István Koren}[%
orcid=000-0003-1350-6732,
email=koren@pads.rwth-aachen.de,
url=https://istvank.eu,
]
\cormark[1]

\address[]{Chair of Process and Data Science,
  RWTH Aachen University, Germany}

\author[]{Jan Niklas Adams}[%
orcid=0000-0001-8954-4925,
email=niklas.adams@pads.rwth-aachen.de,
url=https://niklasadams.com/,
]

\author[]{Alessandro Berti}[%
orcid=0000-0002-3279-4795,
email=a.berti@pads.rwth-aachen.de,
url=https://www.alessandroberti.it/,
]

\author[]{Wil M.P. {van der Aalst}}[%
orcid=0000-0002-0955-6940,
email=wvdaalst@pads.rwth-aachen.de,
url=https://vdaalst.com/,
]

\cortext[1]{Corresponding author.}

\begin{abstract}
Process mining has become a cornerstone of process analysis and improvement over the last few years.
With the widespread adoption of process mining tools and libraries, the limitations of traditional process mining to deal with event data with multiple case identifiers, i.e., object-centric event data, have become apparent.
As a response, the subfield of object-centric process mining has formed, including a file format standardization attempt in the form of OCEL 1.0, unifying the insights of previous developments in capturing object-centric event data.
However, discussions among researchers and practitioners have shown that the proposed OCEL 1.0 standard does not go far enough.
OCEL 2.0 has been proposed as an advanced refinement, including normative and explicit object-to-object relationships, qualifiers for object-to-object and event-to-object relationships, and evolving object attribute values.
This demonstration presents the OCEL 2.0 website available under the URL \url{https://www.ocel-standard.org} as a one-stop shop for the detailed specification, example event logs, and broad tool support to facilitate the adoption of the format.
\end{abstract}


\maketitle

\section{Introduction}
Process mining has permeated many organizations over the last few years.
A wide range of dedicated tools and functionalities within off-the-shelf software systems can perform process analysis.
Although process mining can yield powerful insights, there are currently some limiting factors.
Often, only case-based analysis is offered.
Therefore, for each new perspective on a process or related processes, new extractions and transformations from the information systems have to be performed.
Traditional event logs stored in the XES standard format can only store and provide case-based event data.
Object-centric process mining addresses this limitation~\cite{Aals19,Fahl22}, broadening the horizon of many enterprises to look at intertwined processes within and across their business units and organizations.
The OCEL (1.0) format~\cite{GPBA20} has been initially introduced as a standardization attempt to store and exchange object-centric event data.
However, industry and academia have found that this format is not expressive enough and should be moved closer to reality, e.g., recording object changes and giving detailed information about object relationships~\cite{WLA*21}. These limitations are the motivation for the OCEL 2.0 standard, which refines the original format introducing \emph{object-to-object relationships}, \emph{evolving object attribute values}, and qualifiers for \emph{object-to-object} and \emph{event-to-object relationships} altogether with an increased interpretability of its specifications while maintaining backward compatibility to OCEL 1.0 and graph-based object-centric storage formats.
In this paper, we specifically introduce the OCEL 2.0 website, a one-stop shop for the OCEL 2.0 standard and its resources, featuring the specification of the standard, multiple simulation models, event logs, and tool support to immediately benefit from the new constructs.
This comprehensive set of training materials is collected on the
website \url{https://www.ocel-standard.org}.

\section{Website}
The website serves as an exhaustive resource for both researchers and practitioners.
It provides access to the latest specification along with a scientific formalization of the standard, and a number of event logs.
Additionally, it offers an array of libraries and examples to get started.

\subsection{Specification}
\begin{figure*}[t]
\centerline{
\includegraphics[width=1.0\textwidth]{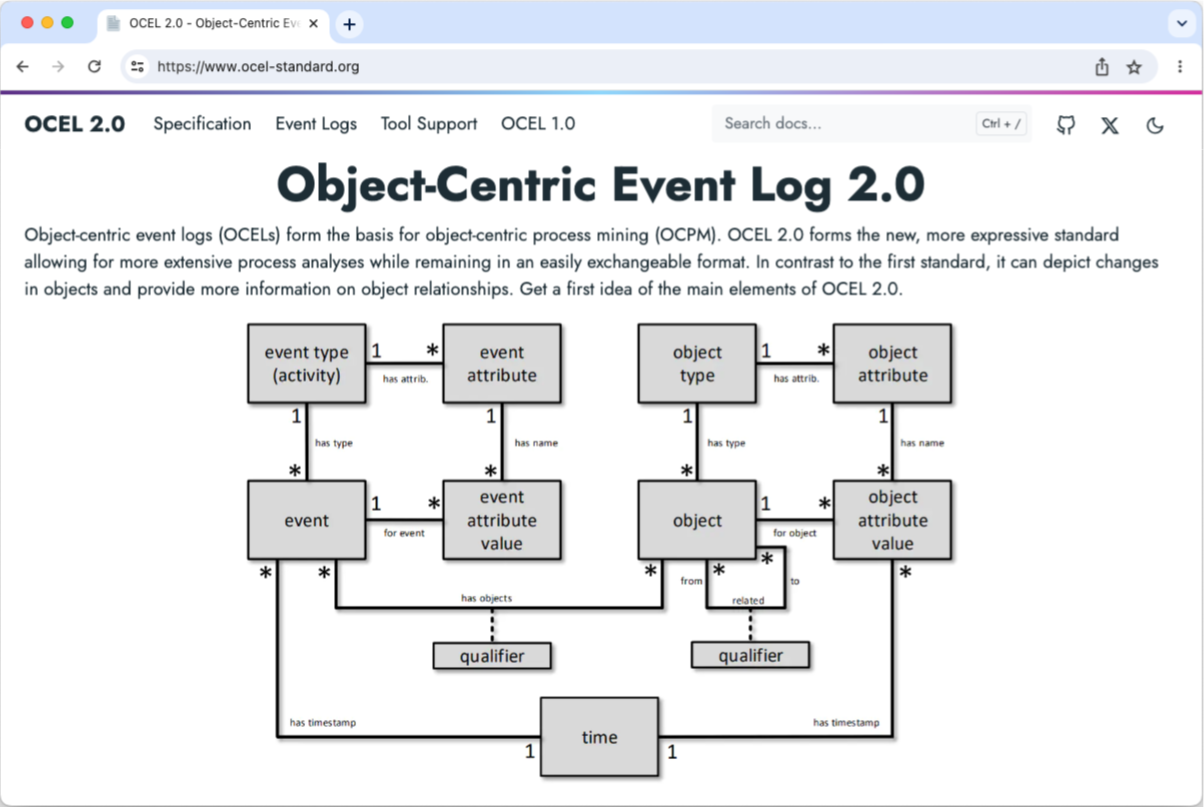}
}
\caption{The OCEL 2.0 website and metamodel.}
\label{fig:website}
\end{figure*}

Figure~\ref{fig:website} shows a screenshot of the website and the OCEL 2.0 metamodel.
The greatest strength of the format is that various relationships and their qualifiers are now directly incorporated.
Events now have a qualified relationship to objects, which means that the nature of the relationship is explicitly annotated.
Similarly, object-to-object relationships can be explicitly stored and supplemented with a qualifier.
Both reinforce the benefit of flexibility by storing various intertwined processes without having to refer back to source data.
In addition, changes to object attribute values can now be tracked.
We provide three file storage formats.
The \emph{XML} variant is human-readable and structured according to industry XML best practices.
The \emph{JSON} variant features a lightweight structure that can be natively used in web environments.
The \emph{SQLite} variant allows for efficient SQL querying, eliminating the necessity to parse the entire event log.

\subsection{Event Logs}
The website initially provides four new event logs that target and demonstrate new features of OCEL 2.0, available on the research data platform Zenodo.

The \textit{Logistics}~\cite{KnGr23} log describes the processes of a company selling goods overseas.
After receiving an order, the shipment of the goods is scheduled.
According to the schedule, the goods are picked up from the local production site and brought to a terminal, where a service provider receives and ships them.

The \textit{Order Management}~\cite{KnAa23} simulation and event log deals with an imaginary e-commerce enterprise.
It comprises both the registration and payment of incoming orders, as well as the process of packing and shipping them.
For these tasks, the company deploys staff in their sales, warehousing, and shipment departments.

The \textit{Procure-to-Pay}~\cite{PaTa23} log is a simulated event log that has been replicated with domain knowledge of a real SAP system.
It handles scenarios such as lengthy approvals and duplicate payments.
Resource availability is realistically mapped based on business hours, and the capacity of human resources.

The \textit{Angular GitHub Commits}~\cite{Pego23} is a real-world event log containing information on the commits in the GitHub repository of the Angular project.
Additionally, we converted some event logs that have already been made available for OCEL 1.0.

\subsection{Tool Support}
Several tools and libraries have been developed to accommodate the OCEL 2.0 format:
\begin{itemize}
\item \emph{OC$\pi$}~\cite{AdAa22}: A standalone tool for object-centric process discovery, variant analysis, and filtering. OC$\pi$ supports object-centric variants and filters based on frequency of behavior across object types.
\item \emph{Object-Centric Process Mining (OCPM)}~\cite{BeAa23}: A JavaScript web application that provides capabilities such as process discovery and visualization of object-centric models.
\item \emph{Ocelot.pm}: A web-based event log inspector for OCEL 2.0 files. It visualizes event and object types as models and shows events and objects in searchable data tables.
\item \emph{Oracle EBS Connector}: A transformer connecting to Oracle databases, targeting tables related to the Purchase-to-Pay process, and facilitating OCEL 2.0 export.
\item \emph{Celonis OCDM Connector}: A tool designed for the Celonis platform, enabling import/export of object-centric event logs in OCEL 2.0 to the Celonis \emph{Object-Centric Data Model}.
\item \textsc{ocpa} \emph{library}~\cite{APAa22}: Focuses on processing object-centric event logs, process discovery, model evaluation, performance analysis, and predictive monitoring.
\item \emph{pm4js library}: For JavaScript environments, it supports tasks like log flattening, statistics generation, process discovery, and more.
\item \emph{pm4py library}~\cite{BZSc23}: A prominent Python library in process mining that has incorporated OCEL 2.0 support, compatible with the XML, JSON, and relational storage formats.
\end{itemize}

\section{Conclusion}
Using object-centric process mining, analysts can get a holistic overview of relationships between different processes and their objects.
The new OCEL 2.0 specification overcomes the limitations of the OCEL 1.0 format by providing qualifiers, object-to-object relationships, and changes to object attribute values.
Our website builds a comprehensive one-stop shop with all the content necessary to get started implementing connectors and custom process mining analyses.
In the future, we plan to extend the website with further content, such as additional event logs and tools/libraries.
We are open to external contributions, particularly those that include real-life event logs.

\begin{acknowledgments}
Funded by the Deutsche Forschungsgemeinschaft (DFG, German Research Foundation) under Germany's Excellence Strategy - EXC-2023 Internet of Production - 390621612.
We also thank the Alexander von Humboldt (AvH) Stiftung for supporting our research. 
\end{acknowledgments}

\bibliography{ocel2resources}

\end{document}